\documentclass[superscriptaddress,longbibliography,amsmath,amssymb,aps,prx,notitlepage,twocolumn,floatfix]{revtex4-1}

\usepackage{graphicx,xcolor}
\usepackage{amsmath,amssymb}

\newcommand{\bs}[1]{\boldsymbol{#1}}

\newcommand{\ie}{{\it i.e.},\ }
\def\eg{\emph{e.g.}\ }

\usepackage{dcolumn}
\usepackage{bbold}

\AtBeginDocument{
    \newwrite\bibnotes
    \def\bibnotesext{Notes.bib}
    \immediate\openout\bibnotes=\jobname\bibnotesext
    \immediate\write\bibnotes{@CONTROL{REVTEX41Control}}
    \immediate\write\bibnotes{@CONTROL{%
    apsrev41Control,author="08",editor="1",pages="1",title="0",year="1"}}
     \if@filesw
     \immediate\write\@auxout{\string\citation{apsrev41Control}}%
    \fi
}

\begin{document}

\title{Dominant Kitaev interactions in the honeycomb materials\\
Na$_3$Co$_2$SbO$_6$ and Na$_2$Co$_2$TeO$_6$}

%\author{\phantom{x}}
%\affiliation{{\color{red}Authors and order of authors to be discussed:}}
%\author{Teams from Melbourne, Chicago, Hamburg, Okayama}

\author{Alaric L.\ Sanders}
\affiliation{School of Physics, University of Melbourne, Parkville, VIC 3010, Australia}

\author{Richard A.\ Mole}
\affiliation{Australian Nuclear Science and Technology Organisation (ANSTO),
New Illawarra Road, Lucas Heights, New South Wale 2234, Australia}

\author{Jiatu Liu}
\affiliation{School of Chemistry, The University of Sydney, Sydney, NSW 2006, Australia}

\author{Alex J.\ Brown}
\affiliation{School of Chemistry, The University of Sydney, Sydney, NSW 2006, Australia}

\author{Dehong Yu}
\affiliation{Australian Nuclear Science and Technology Organisation (ANSTO),
New Illawarra Road, Lucas Heights, New South Wale 2234, Australia}

\author{Chris D.\ Ling}
\affiliation{School of Chemistry, The University of Sydney, Sydney, NSW 2006, Australia}

\author{Stephan Rachel} 
\email{Corresponding author: {\tt stephan.rachel@unimelb.edu.au}}
\affiliation{School of Physics, University of Melbourne, Parkville, VIC 3010, Australia}

\date{\today}

%%%%%%%%%%%%%%%%%%%%%%%%%%%%%%%%%%%%%%%%%%%%%%%%%%%%%%%%%%%%%%%%%%%%%%%%%%%
% MACROS

\newcommand{\ncto}{Na$_2$Co$_2$TeO$_6$}
\newcommand{\ncso}{Na$_3$Co$_2$SbO$_6$}

%%%%%%%%%%%%%%%%%%%%%%%%%%%%%%%%%%%%%%%%%%%%%%%%%%%%%%%%%%%

\begin{abstract}
Cobaltates with 3$d$ based layered honeycomb structure were recently proposed as Kitaev magnets and
putative candidates to host the long-sought Kitaev spin liquid. Here we present inelastic neutron
scattering results down to 50 mK for powder samples of Na$_3$Co$_2$SbO$_6$ and Na$_2$Co$_2$TeO$_6$,
with high resolution in regions of low momentum and energy transfers. We
compare the experimental data below the antiferromagnetic zigzag ordering temperature with dynamical
structure factors obtained within spin wave theory. We search the wide parameter range of a
$K$-$J_1$-$\Gamma$-$\Gamma^{\prime}$-$J_3$ spin 1/2 model and identify the best fits to constant
momentum cuts of the inelastic neutron data. The powder average limits our ability to uniquely
select a best-fit model, but we find that the experimental data is matched equally
well by two classes of parameters:
one with a dominant $K<0$, $|K/J_1|\sim 5\ldots 25$, and another
with $K>0$, $|K/J_1| \sim 1$. We show
that these classes are equivalent under the exact self-duality
transformation identified by Chaloupka and Khalliulin \cite{chaloupka_hidden_2015}.
This model symmetry unifies a number of previous parameter estimates.
Though the two cases are indistinguishable by our experiment,
there is evidence in favour of the $K<0$ case.
A purely isotropic Heisenberg model is incompatible with our results.

\end{abstract}

\maketitle

%%%%%%%%%%%%%%%%%%%%%%%%%%%%%%%%%%%%%%%%%%%%%%%%%%%%%%%%%%%

%%%%%%%%%%%%%%%%%%%%%%%%%%%%%%%%%%%%%%%%%%%%%%%%%
%
%                                                            I N T R O
%
%%%%%%%%%%%%%%%%%%%%%%%%%%%%%%%%%%%%%%%%%%%%%%%%%
\section{Introduction}
\label{sec:intro}

The search for quantum spin liquids (QSLs) is one of the great challenges in the field of
strongly correlated
electrons\,\cite{anderson73mrb153,fazekas-74pm423,balents-10n199,savary-17rpp016502,zhou-17rmp025003}.
A QSL state possesses short-ranged magnetic fluctuations but no long-range order. It is, hence, a
featureless state which is mostly characterized by the absence of clear and easily-identifiable
features. Despite many years of extensive research, few candidate spin liquid materials are
known\,\cite{helton-07prl107204,han-12n406,powell-11rpp056501}, and the list of theory proposals
for experimental realisations is rather short. Mott insulators with geometric or spin-exchange frustration are the traditional places to search for QSLs\,\cite{balents-10n199}. As an additional complication, in two and three spatial dimensions models for QSLs can usually not be solved exactly and theoreticians rely on approximate numerical methods to identify and characterize such states.

Kitaev's seminal work opened an alternative avenue, where he introduced an exactly soluble QSL Hamiltonian with bond-dependent Ising interactions on the honeycomb lattice\,\cite{kitaev_anyons_2006}. The exact solution revealed that the QSL consists of free Majorana fermions coupled to a static $\mathbb{Z}_2$ gauge field, nicely illustrating the fractionalization of the original spin degrees of freedom. 

This purely theoretical development was further fueled by the influential work by Jackeli and
Khaliullin\,\cite{jackeli_mott_2009} who proposed that Kitaev model might be
realized in transition metal oxides with strong spin-orbit coupling. The idea is based on an observation that the exchange interactions between spin-orbital entangled moments are highly anisotropic and also depend on the bond directions\,\cite{khaliullin05ptss155}. 
For a pedagogical review about spin-orbit entangled states of matter see Ref.\,\cite{takayama-21jpsj062001}.
The first Kitaev candidate
material was Na$_2$IrO$_3$\,\cite{singh-10prb064412,liu-11prb220403} along with its sister compound
$\alpha$-Li$_2$IrO$_3$\,\cite{singh-12prl127203}, which both turned out to be magnetically ordered
at low temperatures. Despite this setback, subsequent work substantiated the claim that these
iridates possess non-negligible Kitaev spin interactions along with other generic spin
exchange\,\cite{choi-12prl127204,ye-12prb180403,mazin-12prl197201,price-12prl187201,comin-12prl266406,reuther-12prb155127,gretarsson-13prl076402,gretarsson-13prb220407,foyevtsova-13prb035107,reuther-14prb100405,kim-14prb081109,rau-14prl077204,chun-15np462,kimchi-15prb245134,li-15prb161101,chaloupka_hidden_2015,rousochatzakis-15prx041035,nishimoto-16nc10273,catuneanu-16prb121118,williams-16prb195158,laubach-17prb121110,yadav-18prb121107,baez-19prb184436, takagiConceptRealizationKitaev2019}.

The second wave of experiments focussed on $\alpha$-RuCl$_3$. While its low-temperature phase revealed magnetic long-range order of the zigzag type\,\cite{sears-15prb144420} just like Na$_2$IrO$_3$, its excitation spectrum measured with inelastic neutrons was interpreted as a combination of magnons stemming from the magnetic order and additional features due to the system's proximity to a QSL\,\cite{banerjee_proximate_2016,sizyuk-16prb085109,wolter-17prb041405,yadav-18prb121107,lampen-kelley-18prb100403}. 
A milestone was the report of a quantized thermal conductance when a magnetic field is applied to $\alpha$-RuCl$_3$\,\cite{kasahara-18n227,yokoi-21s568}: the measured conductance plateau is in agreement with a chiral Majorana mode as expected from Kitaev's model with applied magnetic field (corresponding to a non-Abelian QSL)\,\cite{kitaev_anyons_2006}. It is generally thought that the generic spin exchange spoiling the QSL state in zero field needs to be suppressed by a sufficiently strong magnetic field, until the remaining compass interactions can dominate and cause the non-Abelian QSL state to prevail.

A major challenge in obtaining a better theoretical understanding is to identify the details of the
generic {\it non-Kitaev} interactions in the candidate materials. However, not even the strength  and
sign of the Kitaev compass interactions are known with certainty. 
Experimental determination of these coupling constants is necessarily indirect. Previous estimates
have come from magnetic susceptibility measurements
\cite{vivancoCompetingAntiferromagneticferromagneticStates2020, dasMagneticAnisotropyAlkali2019, lampen-kelley-18prb100403}, Raman
spectroscopy \cite{guptaRamanSignaturesStrong2016}, X-ray scattering
\cite{searsFerromagneticKitaevInteraction2020,
hwanchunDirectEvidenceDominant2015,suzuki-21nc4512} and most commonly, inelastic neutron scattering
\cite{choi-12prl127204,banerjeeNeutronScatteringProximate2017,banerjee_proximate_2016,songvilay-20prb224429,park-20arXiv2012.06167,samarakoon-21arXiv2105.06549}. Theoretical determination of these coupling
constants is a very subtle task, ultimately relying on density functional theory
\cite{janssenMagnetizationProcessesZigzag2017,sizyuk-16prb085109,sizyukImportanceAnisotropicExchange2014,winterChallengesDesignKitaev2016,foyevtsova-13prb035107,li-15prb161101,
kimSpinorbitalEntangledState2021, 
bhattacharjeeSpinOrbitalLockingEmergent2012,rauSpinOrbitPhysicsGiving2016,liu_pseudospin_2018}.

While most attention was given to the iridates and ruthenates of d$^5$ ions Ir$^{4+}$ and Ru$^{3+}$,
respectively, a recent proposal emphasized that cobaltates of d$^7$ ions such as Co$^{2+}$ might be
another platform for Kitaev candidate materials with
pseudospin-$\frac{1}{2}$\,\cite{liu_pseudospin_2018,liu-20prl047201,sano_kitaev-heisenberg_2018},
despite some skepticism that spin-orbit coupling is insufficient to promote compass interactions.
The first series of experimental results including \ncto\ and \ncso\ confirmed that their
low-temperature phase is again antiferromagnetically zigzag-ordered\,\cite{wong_zig-zag_2016,
lefrancois_magnetic_2016, bera_zigzag_2017}. Spin-wave theory modelling of neutron
scattering data has suggested that Heisenberg-Kitaev type models can capture the low energy physics
in the materials\,
\cite{park-20arXiv2012.06167, lin_field-induced_2021, songvilay-20prb224429,
samarakoon-21arXiv2105.06549, liu_towards_2021, kimSpinorbitalEntangledState2021}, however the
community is yet to reach a
consensus as to the sign and magnitude 
of the Kitaev interaction (see Tab. \ref{tab:model_params}).

Here we present inelastic neutron
scattering results for Na$_3$Co$_2$SbO$_6$ and Na$_2$Co$_2$TeO$_6$. In particular, our data was measured at temperatures as low as 50 mK 
with high resolution in regions of low momentum and energy transfers. We model the dynamical
structure factor within spin wave theory, and show that a purely isotropic $J_1$--$J_2$--$J_3$ model
is {\it incompatible} with the experimental data. By fitting our model to our experimental data we
show that best fits are obtained for extended Heisenberg--Kitaev models with either ferro- or
antiferromagnetic Kitaev exchange $K$. 
Good quality data is available
in regions of large momentum and energy transfer \cite{park-20arXiv2012.06167, lin_field-induced_2021, songvilay-20prb224429,
samarakoon-21arXiv2105.06549}, which have established that Kitaev type models are capable of
capturing a majority of features seen in the powder averaged inelastic spectrum.

We instead concentrate on the excitations close to zero energy. This fine
	detail allows us to perform an accurate indirect measurement of the energy gap, which in turn
	establishes a constraint on the values
of the symmetry breaking $\Gamma, \Gamma^\prime$ terms.
We argue that the most likely scenario for both materials is
a dominant ferromagnetic $K<0$ with $|K/J_1|$ being in the range $5 - 25$. Hence, our results indicate that perturbing the magnetically ordered ground state by strain\,\cite{rachel-16prl167201,perrault-17prb184429}, pressure\,\cite{yadav-18prb121107} or an applied magnetic field\,\cite{janssenMagnetizationProcessesZigzag2017,kasahara-18n227} would be an exciting attempt to push the material into the QSL phase.

The paper is organised as follows. In Sec.\,\ref{sec:sample-prep} we discuss the sample preparation of the materials used in Sec.\,\ref{sec:INS} to perform inelastic neutron scattering experiments. In Sec.\,\ref{sec:SWT} we perform linear spin wave theory and derive the scattering intensity which we fit to the experimental data in Sec.\,\ref{sec:fit}. In the discussion in Sec.\,\ref{sec:discussion}, we elaborate on a model duality between ferromagnetic and antiferromagnetic $K$ in agreement with our best fits. Next, we argue that both excitation spectra are gapped, albeit with a much smaller gap for \ncto\ as previously reported. We discuss the classical ground states and the corresponding spin quantization axis, and elaborate on predictions for single crystal measurements. The paper ends with a conclusion in Sec.\,\ref{sec:conclusion}.

%%%%%%%%%%%%%%%%%%%%%%%%%%%%%%%%%%%%%%%%%%%%%%%%%
%
%                                      S A M P L E   P R E P A R A T I O N
%
%%%%%%%%%%%%%%%%%%%%%%%%%%%%%%%%%%%%%%%%%%%%%%%%%
\section{Sample preparation}
\label{sec:sample-prep}
 
The same polycrystalline sample of \ncso\ was used in this work as in Wong \textit{et al.}\,\cite{wong_zig-zag_2016}. A polycrystalline sample of \ncto\ was prepared from Na$_2$CO$_3$ (Merck,
99.9\%), Co$_3$O$_4$ (Sigma-Aldrich, 99.99\%), and TeO$_2$ (Sigma-Aldrich, 99.9995\%). Following the
method described by Berthelot \textit{et al.}\,\cite{berthelotStudiesSolidSolutions2012}, the
reagents were ground together at the correct stoichiometry, pressed into pellets, and calcined twice
at 850$^\circ$C in air for 12 hours with intermediate regrinding. Sample purity was confirmed by
X-ray powder diffraction. Magnetic property measurements were also consistent with previous reports
for \ncto: zero-field cooled temperature-dependent susceptibility in a 0.1\,T field showed a sharp
antiferromagnetic transition at $T_N=27$\,K and a weaker feature at 16\,K,  which were suppressed under field-cooled
conditions. A Curie-Weiss fit to the paramagnetic region yielded an effective magnetic moment of
4.02 $\mu_B$/Co (within the range typically observed for Co$^{2+}$) and a Weiss constant
$\theta_{\rm CW} = -1.4$\,K.

\section{Inelastic Neutron Scattering}
\label{sec:INS}

Inelastic neutron scattering data were collected using the cold-neutron time-of-flight spectrometer
Pelican\,\cite{yu_pelican_2013, yu_performance_2015} at the Australian Centre for Neutron Scattering.
Approximately 5\,g
of each sample was held in an annular sample can fabricated from oxygen-free copper.  This was
cooled using a dilution insert inside a top-loading closed cycle cryostat.  The instrument was
aligned for 4.69\,\AA\ neutrons. The choppers were also rephased to allow the collection of data
with $\lambda/2 = 2.345$\,\AA\ neutrons. Data were collected at 50\,mK and 1.8\,K and
corrected for background by subtraction of an empty can and normalised to a
standard vanadium sample. All raw detector data were processed using the freely available
LAMP\,\cite{richard_analysis_1996} software.

%\begin{figure*}[h!]
%	\centering
%	\includegraphics[width=\textwidth]{fig/raw_6fig_2.png}
%
%	\caption{Inelastic neutron scattering results at 0.05K, $\sim1.7$K and 7.4K for powder
%		samples of \ncto (a-c) and \ncso (d-f). Neutron scattering at $\lambda = 2.345$\AA are
%		shown in panels (a) and (d); remaining plots were collected at $\lambda=4.69$\AA. \label{fig:ins_data}
%	}
%\end{figure*}
%
%%%%%%%%%%%%%%%%%%%%%%%%%%%%%%%%%%%%%%%%%%%%%%%%%%%%%%%
\begin{figure}[t!]
	\centering
	\includegraphics[width=\columnwidth]{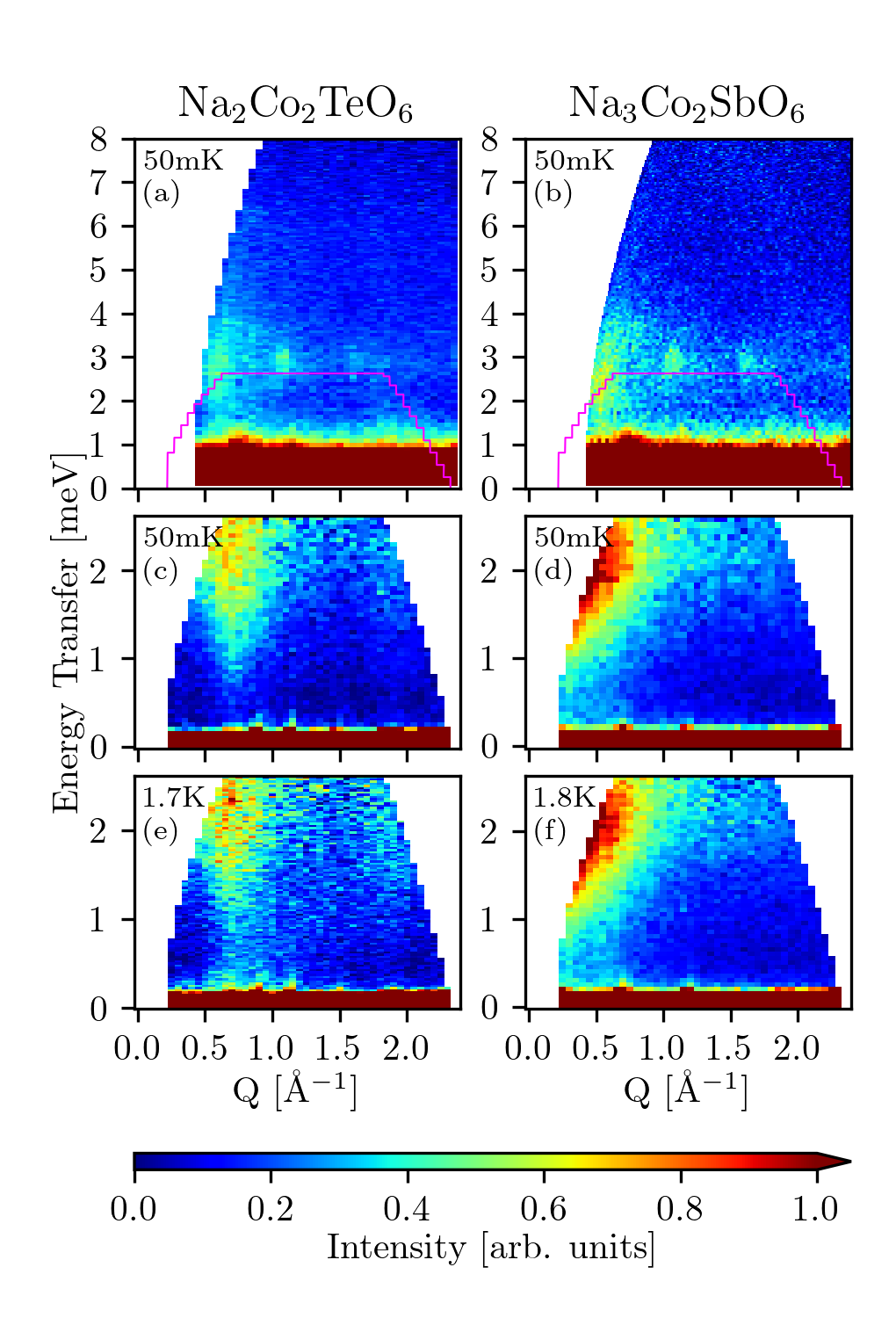}

	\caption{
	Inelastic neutron scattering results for powder
		samples of \ncto\ (left column) and \ncso\ (right column). (a-d) are measured at $T=50$\,mK
		while (e) and (f) at 1.7\,K and 1.8\,K, respectively. Detector resolution is 0.135$\mu$eV,
		0.01 \AA$^{-1}$.
		Data gathered at
		$\lambda/2 = 2.345$ \AA\ are
		shown in panels (a) and (b), remaining plots were collected at
		$\lambda=4.69$\,\AA.
		Pink curves on (a) and (b) correspond to domain of intensity data shown in (c-f). 
		\label{fig:ins_data}
	}
\end{figure}
%%%%%%%%%%%%%%%%%%%%%%%%%%%%%%%%%%%%%%%%%%%%%%%%%%%%%%%

Magnetic Bragg reflections were observed at $Q=0.75$\AA, confirming that the system was indeed
magnetically ordered for the measured temperatures.
It is common to observe up to three phase transitions in \ncto\ - a well-known transition at $\sim$26K coinciding
with the appearance of zigzag magnetic Bragg peaks \cite{bera_zigzag_2017}, a weaker
transition at 31K \cite{chen-21prbL180404}, and the weakest at 14K
\cite{yaoFerrimagnetismAnisotropicPhase2020}. The 14K transition may be due to a subtle
magnetic effect, such as interlayer ordering. Ref. \cite{samarakoon-21arXiv2105.06549} includes a
weak interlayer Heisenberg coupling when calculating spin-wave dispersions, from which one may
estimate its magnitude at around $\sim 0.4$meV. We concern ourselves principally with the low-lying excitations of
the high-symmetry $M$ point, at which the dynamics are dominated by the SO(3) symmetry breaking
terms responsible for opening the gap.

\ncto\ at 50\,mK clearly features a
mode emanating from $Q=0.75$ \AA, corresponding to the $M$ point in the Brillouin zone of the honeycomb layers.
The dispersion relation close to this point appears to be linear. Our results are consistent with
existing experiments \cite{park-20arXiv2012.06167, park-20arXiv2012.06167,songvilay-20prb224429,
chen-21prbL180404, samarakoon-21arXiv2105.06549}, however, our ability to resolve features below 1\,meV
provides crucial detail at the gap point (\ie where the magnon excitation gets closest to the elastic line). The single-$Q$ cuts of the data in Fig.\,\ref{fig:ins_data}\,(c), shown in
Fig.\,\ref{fig:all_cuts}\,(a,c), exhibit a statistically significant excess of spectral weight down to the
elastic line with an essentially linear intensity fall-off. This seemingly gapless spectrum 
disagrees somewhat with existing models, which are generally gapped by several meV. 
It it not obvious
\textit{a priori} how large of a modification to these models is necessary to fit our experiment.

The spectrum for \ncso\ shows no evidence of any low-energy modes near the $M$ or $K$ points, the sole resolvable magnon excitation 
appears to emanate from $\Gamma$. The dispersion resembles a quadratic mode, which may be compatible
with the magnon spectrum predicted by Ref.\,\cite{liu-20prl047201}. Unlike \ncto, there is very little
difference between the two temperatures in panels \ref{fig:ins_data}\,(d) and (f).

%%%%%%%%%%%%%%%%%%%%%%%%%%%%%%%%%%%%%%%%%%%%%%%%%
%
%                                                S W T
%
%%%%%%%%%%%%%%%%%%%%%%%%%%%%%%%%%%%%%%%%%%%%%%%%%
\section{Spin Wave Theory}
\label{sec:SWT}

\begin{figure}
	\includegraphics[width=\columnwidth]{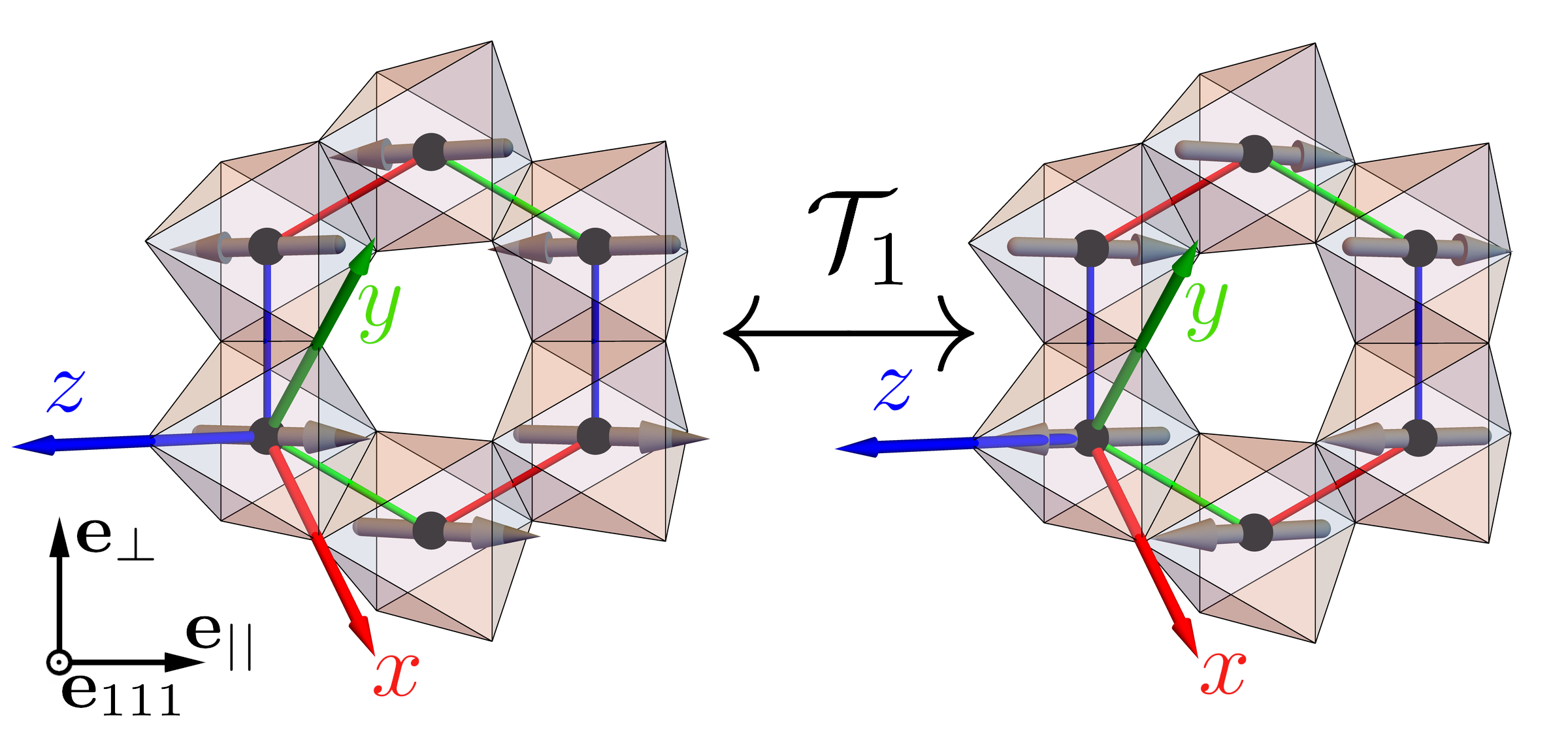}
\caption{
        Classical magnetic moments (gray) of $j=\frac{1}{2}$ pseudospins relative to CoO$_8$
	octahedra (peach). Left and right diagrams show conjugate models under the duality
	transformation $\mathcal{T}_1$. Red, green and blue bonds correspond to Kitaev interactions
along the $x$, $y$ and $z$ axes, respectively. $\mathbf{e}_\|, \mathbf{e}_\perp$ and
$\mathbf{e}_{111}$ are defined in Sec.\,\ref{sec:cgs}.}	\label{fig:3D}
\end{figure}

We model the system using a six-parameter extended Heisenberg-Kitaev model as previously used in Refs.\,\cite{songvilay-20prb224429, park-20arXiv2012.06167, park-20arXiv2012.06167,
samarakoon-21arXiv2105.06549}. The first order anisotropic interactions are the most general
possible couplings that preserve $C_3$ symmetry, parameterised by $J_1, K, \Gamma, \Gamma^\prime$ (explained below).
These are supplemented by second and third-nearest neighbour Heisenberg
couplings of respective strengths $J_2$ and $J_3$. This is motivated by \textit{ab intio} calculations
of realistic, distorted octahedral lattices\,\cite{liu-20prl047201} and the failure of
simpler models to accurately reproduce analogous systems such as $\alpha-$RuCl$_3$
and A$_2$IrO$_3$\,\cite{kimSpinorbitalEntangledState2021} (A=Na or Li). 
Following the original proposal for the realisation of bond-anisotropic couplings\,\cite{jackeli_mott_2009},
we take the quantisation axes $x, y, z$ to
be oriented as in Fig.\,\ref{fig:3D}, \ie normal to the corresponding edge-sharing rectangles.
The Hamiltonian may be expressed in the form 
\begin{align}
	\mathcal{H} &= \sum_{\langle ij \rangle^\gamma}\Big\{ J_1
		\mathbf{S}_i\cdot\mathbf{S}_j  + K S_i^\gamma S_j^\gamma  + \Gamma
	\left(S_i^\alpha S_j^\beta + S_i^\beta S_j^\alpha \right)\nonumber \\[2pt]
		    &+ \Gamma^{\prime}\left(S_i^\gamma
S_j^\alpha + S_i^\gamma S_j^\beta + S_i^\alpha S_j^\gamma + S_i^\beta S_j^\gamma \right) \Big\} \label{ham}
 \\[10pt]
%	\mathcal{H} &= \sum_{\langle ij \rangle_1^\gamma} \mathbf{S}_i^T \left(K_\gamma +
%	J_1 I\right) \mathbf{S}_j
	\nonumber    &+ J_2\sum_{\langle\!\langle ij \rangle\!\rangle} \mathbf{S}_i\cdot
	\mathbf{S}_j
	+ J_3\sum_{\langle\!\langle\!\langle ij \rangle\!\rangle\!\rangle} \mathbf{S}_i\cdot \mathbf{S}_j\ .
\end{align}
Here $J_i$ represents $i$th neighbor isotropic Heisenberg spin exchange, $K$ is nearest-neighbor Kitaev coupling while $\Gamma$ and $\Gamma^{\prime}$ are symmetric off-diagonal couplings. Indices within single (double and triple) brackets $\langle \cdot \rangle$ denote nearest (2nd and 3rd nearest) neighbors lattice sites, and $\langle ij \rangle^\gamma$ is a $\gamma$ bond with $\gamma=x, y, z$. On $z$ bonds we assume $(\alpha, \beta, \gamma)=(x,y,z)$ with cyclic permutations on $x$ and $y$ bonds. The first two lines of Eq.\,\eqref{ham} may be written more compactly as $\mathcal{H}^{(1)} = \sum_{\langle ij \rangle^\gamma} \mathbf{S}_i^T K_\gamma \mathbf{S}_j $
using the matrices 
\begin{align}\label{Kxy}
	K_x = \begin{pmatrix} 
		K+J & \Gamma^{\prime} & \Gamma^{\prime}\\
		\Gamma^{\prime} & J & \Gamma \\ 
		\Gamma^{\prime} & \Gamma & J
	\end{pmatrix}&, ~~
	K_y = \begin{pmatrix} 
		J & \Gamma^{\prime} & \Gamma\\
		\Gamma^{\prime} & K+J & \Gamma^{\prime} \\ 
		\Gamma & \Gamma^{\prime} & J
	\end{pmatrix}, \\[5pt]   \label{Kz}
	K_z =& \begin{pmatrix} 
		J & \Gamma & \Gamma^{\prime} \\
		\Gamma & J & \Gamma^{\prime} \\
		\Gamma^{\prime} & \Gamma^{\prime} & K+J
		\end{pmatrix}\ .
\end{align}

%\begin{align}
%	\mathcal{H} &= \sum_{\langle ij \rangle_1^\gamma} \mathbf{S}_i^T K_\gamma \mathbf{S}_j 
%		    &+ J_2\sum_{\langle ij \rangle_2} \mathbf{S}_i\cdot \mathbf{S}_j
%	+ J_3\sum_{\langle ij \rangle_3} \mathbf{S}_i\cdot \mathbf{S}_j
%\end{align}

%%%%%%%%%%%%%%%%%%%%%%%%%%%%%%%%%%%%%%%%%%%%%%%%%
%%%%%%%%%%%%%%%%%%%%%%%%%%%%%%%%%%%%%%%%%%%%%%%%%

The magnetic Bragg peaks observed in both materials at our experiment are consistent with the results of previous work\,\cite{bera_zigzag_2017, wong_zig-zag_2016, lefrancois_magnetic_2016}, which suggest 
a zig-zag ground state. We employ standard linear spin-wave theory to
probe the quantum fluctuations about a polarised classical ground state. The
ground state vector is not chosen spontaneously -- in general, the anisotropic interactions break
SU(2) symmetry fixing the orientation of the magnetic moments relative to the lattice. This optimal
ground state is found analytically in section \ref{sec:cgs}.

About this ground state, we expand in Holstein-Primakoff bosons, from which we deduce the magnon
dispersion $\epsilon_i(\mathbf{Q})$ and spectral weight function $S^{\alpha\beta}(\mathbf{Q},\omega)$. In the linearised
approximation, all three-operator terms are discarded and excited states have infinite lifetimes,
resulting in $S(\mathbf{Q},\omega) \propto \sum_{i=1}^N
S_i(\mathbf{Q})\delta(\omega-\epsilon_i(\mathbf{Q})) $,
where $N=4$ is the number of magnetic sites per unit cell. 

We capture the effects of finite detector resolution and
magnon damping by convolving the spectrum with a Voigt profile. The Gaussian standard deviation
introduced by the settings of the Pelican spectrometer is $0.135$\,meV, but the Lorentzian broadening was determined
phenomenologically. We fit a slice from the single crystal data of Ref.\,\cite{chen-21prbL180404}, finding $\gamma
= 0.25(5)$\,meV to give an acceptable fit to the low-energy peaks (see Fig.\,\ref{Sfig:fits} in
App.\,\ref{sec:appA}). Although the broadening is in general $Q$-dependent, the single crystal evidence seems to
suggest that any variation is fairly weak\,\cite{chen-21prbL180404}.

It has recently been suggested \cite{maksimovRethinkingEnsuremathAlpha2020} that corrections from
many-magnon interactions are essential when fitting frustrated Kitaev-like models to high-energy
($>$3meV) magnon spectra. Magnon energy renormalisations and lifetime broadening can be both
substantial (of order $\sim$1meV) and highly anistropic \cite{smitMagnonDampingZigzag2020,
winterBreakdownMagnonsStrongly2017}, particularly in the presence of large off-diagonal couplings
$\Gamma, \Gamma^{\prime}$. Multi-magnon excited state energies are bounded below by a sharp
dispersive cutoff, corresponding to the minimum combined
energy of two magnons with $k_1 + k_2 = k$. In a neutron scatteirng experiment, this manifests as a sharp change in
broadening when the decay channel $k\to k_1 + k_2$ becomes kinematically allowed
\cite{winterBreakdownMagnonsStrongly2017}. Known single crystal data for the minimum energy mode \cite{chen-21prbL180404} shows
no such broadening discontinuities, suggesting that the linear approximation is appropriate for the
lowest energy mode.

%%%%%%%%%%%%%%%%%%%%%%%%%%%%%%%%%%%%%%%%%%%%%%%%%%%
% 
%                                                               F I T T I N G
%
%%%%%%%%%%%%%%%%%%%%%%%%%%%%%%%%%%%%%%%%%%%%%%%%%%%
\section{Fitting the Powder Spectrum}
\label{sec:fit}

For both materials, our long-wavelength neutron experiments have access to the low energy $E < 4$\,meV
region of $Q$--$E$ space, at the cost of only seeing low-energy details. As we do not have access to the high-energy features that aided the
qualitative fitting of models in Refs.\,\cite{park-20arXiv2012.06167} and \cite{songvilay-20prb224429},
we require a quantitative goodness-of-fit measure to navigate the
5D parameter space. We therefore adopt the approach of Ref.\,\cite{samarakoon-21arXiv2105.06549}.
We define 
\begin{equation}
	\label{eq:chi2}
\chi^2 = \sum_{Q, \omega} \frac{(\mathcal{I}_\text{exp}(Q,\omega) -
\mathcal{I}_\text{theory}(Q,\omega))^2}{\mathcal{I}_\text{exp}(Q,\omega)}
\end{equation}
where $\mathcal{I}_\text{exp}$ is the experimental scattering intensity and
$\mathcal{I}_\text{theory}(Q,\omega) \propto F(Q)^2 S^\perp(Q,\omega)$. $F(Q)$ is the
spherically symmetric atomic form factor of Co$^{2+}$ \cite{watsonHartreeFockAtomic1961}, and
$S^\perp$ is as defined in (\ref{eq:sperp}). The theoretical intensity
function is normalised to have the same total intensity as the experiment, when restricted to pixels
within kinematic limits and at an energy above 4\,meV. This energy cut was used to exclude the elastic
line. All plots shown account for the form factor fall-off of the magnetic Co ions.

Due to the lack of clearly resolved high energy features in our 4.69 \AA\  data on either substance [see Fig.\,\ref{fig:ins_data}\,(a) and (b)], fine detail at low $Q$ provides the most useful information for model fitting. We
therefore used the experimental intensity at 50\,mK from Figs.\,\ref{fig:ins_data}\,(c) and
\ref{fig:ins_data}\,(d) for $\mathcal{I}_\text{exp}$. The domain of the low-energy data is
outlined in pink on Fig.\,\ref{fig:pavg}. 

The powder averaging of the inelastic neutron data from the polycrystalline sample makes unique determination of best-fit parameters
difficult.
Previous reports on these compounds have
consistently found weak $J_2$ when it is included in \ncto\,\cite{samarakoon-21arXiv2105.06549,
park-20arXiv2012.06167,songvilay-20prb224429}, and all previous fit attempts have neglected it when
fitting \ncso. Due to the small region of data available, we set $J_2=0$ to mitigate over-parameterisation.

Model parameters were optimised by the following procedure:
\begin{enumerate}
	\item Suggest a starting guess for the six parameters.
	\item Compute $\chi^2$ for a rectangle of side 2 meV centred on the guess in
		$\Gamma-\Gamma^\prime$ space, $J_1-K$ space and $J_1-J_3$ space.
	\item Compare the local minima of these three phase portraits, and update the guess to
		match.
\end{enumerate}

\begin{figure}[t!]
\centering
\includegraphics[width=\columnwidth]{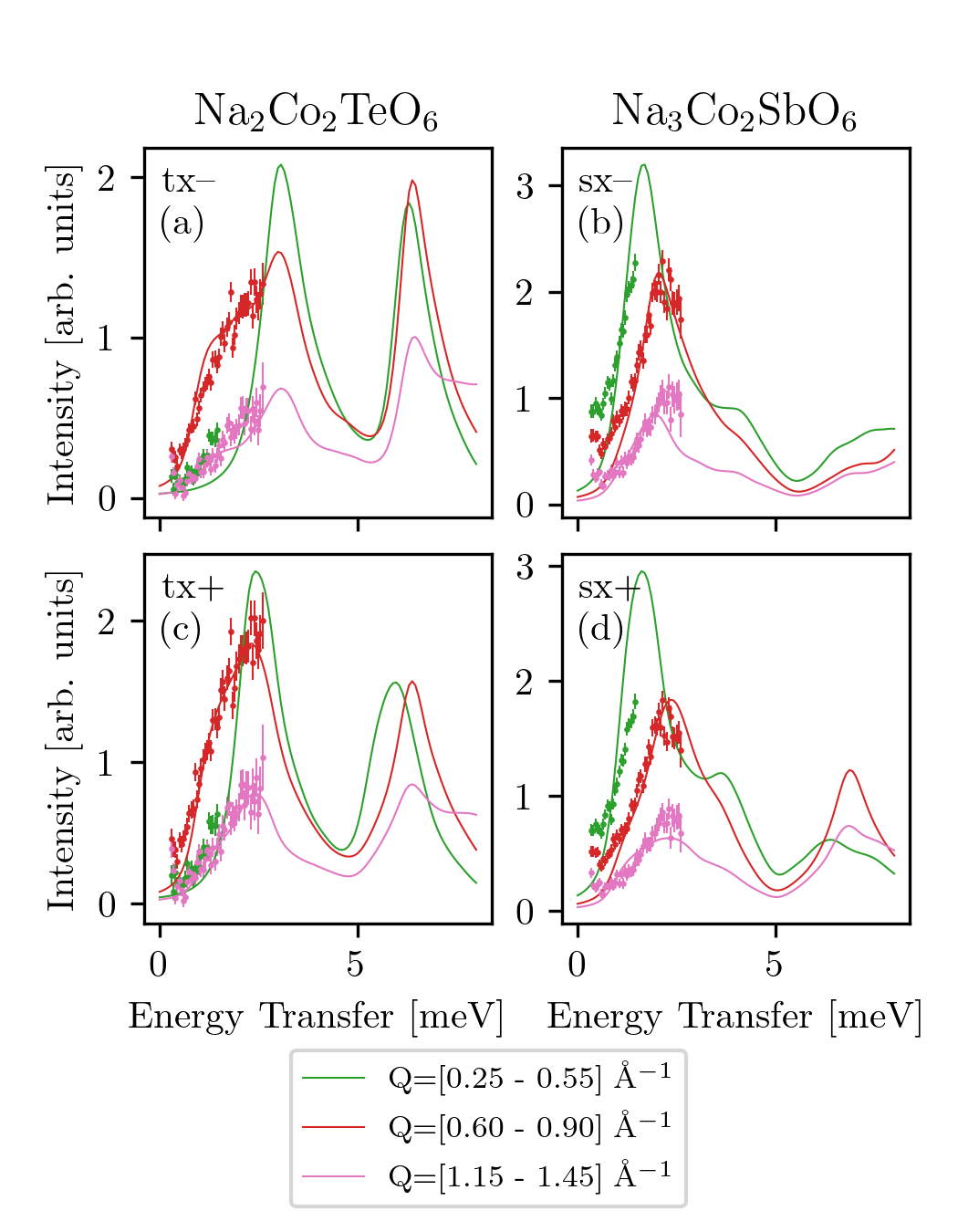}
\caption{Best fit of LSWT models with $K>0$ to the INS data of \ncto\ and \ncso, showing both $K>0$ and
$K<0$ fits.}
\label{fig:all_cuts}
\end{figure}

\begin{figure}[h!]
	\centering
	\includegraphics[width=\columnwidth]{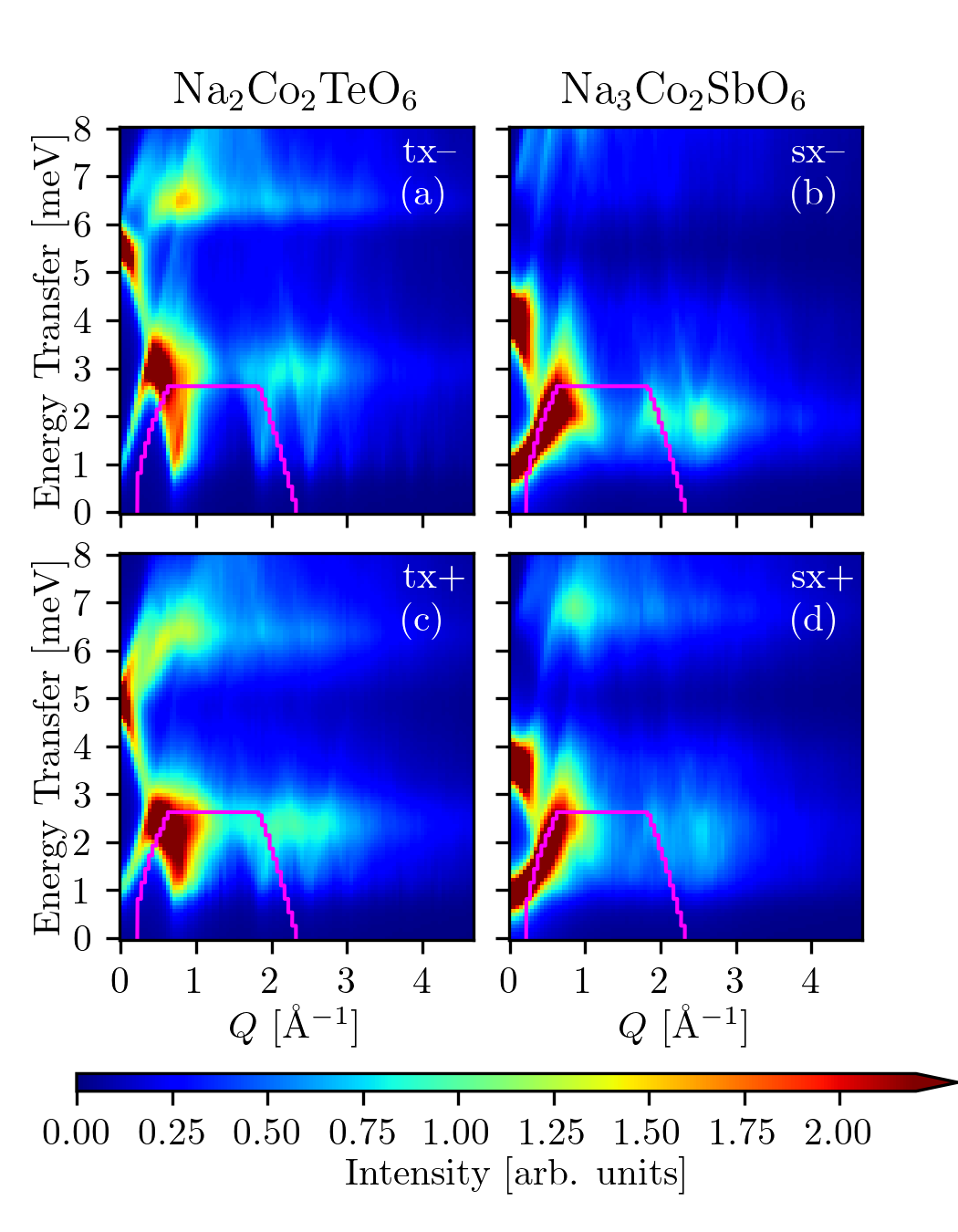}
	\caption{Calculated powder averages for \ncto\ (left column) and \ncso\ (right column).
		\label{fig:pavg}
	}
\end{figure}

This procedure was successful in obtaining close
matches of the binned powder data to the theoretical model. The \ncto\ models in Fig.
\ref{fig:all_cuts}(a),(c) are close matches with the expeirmental cuts, subject to some slight
variations that may be attributed to momentum-space broadening from defects; a factor not accounted
for in the theoretical model. In particular, the low-momentum cut between 0.25\AA$^{-1}$ and
0.55\AA$^{-1}$ is likely contaminated by
diffraction from the zero-order peak - the `hourglass' like signal at Q=0.5\AA$^{-1}$ is consistent with a
broadened central maximum, but inconsistent with a magnetic excitation.

The parameters ultimately obtained should not be interpreted as rigorous global minima of the
error term. Rather, they should be taken as estimates that reflect the general
hierarchy of coupling strengths. It is not possible to assign meaningful error bars to these
parameters - as illustrated in Fig. \ref{fig:gap}, the $\chi^2$ surface is riddled with
sharp valleys, flat plateaus and local minima that are resistant to purely numerical optimisation methods. In practice, it was necessary
to manually choose a minimum. The parameters are
interrelated in a highly non-trivial manner, so any statement about parameter confidence must be
specified as a manifold embedded in the 5D parameter space. The intersection of the valleys given by
our data with the more localised maxima found in Ref.
\cite{samarakoon-21arXiv2105.06549}, serves to specify the parameter ranges more tightly.

We chose the starting guesses for \ncto\ using the work of
Refs.\,\cite{samarakoon-21arXiv2105.06549}, which  
established regions of parameter space in agreement
with their good quality high energy data. We seek to further restrict the allowed parameter space
and use these models to deduce qualitative features of the magnon spectrum, \eg, whether or not it
has a gap in the excitation spectrum. No such study exists for \ncso, so starting guesses were chosen centered on the sa and
sb$\pm$ best-fit parameters from Refs. 
\cite{park-20arXiv2012.06167, songvilay-20prb224429}, which are listed in Tab.\,\ref{tab:model_params}.

%%%%%%%%%%%%%%%%%%%%%%%%%%%%%%%%%%%%%%%%%%%%%%%%%%%%%%%%%%%

%%%%%%%%%%%%%%%%%%%%%%%%%%%%%%%%%%%%%%%%%%%%%%%%%%%%%%%%%%%

\begingroup
%\squeezetable

\begin{table}
%% CoTe
%\begin{tabular}{|c|l|l|l|l|l|l|l|}
%\begin{ruledtabular}
\begin{tabular}{ll|dddddd}
       \hline\hline
       \multicolumn{2}{c|}{\ncto} & J &  K & \Gamma & \Gamma^{\prime} & J_2 & J_3\\
       \hline
       %Songvilay et al. 
       ta-- &\cite{songvilay-20prb224429} & -0.1 & -9 & 1.8 & 0.3 & 0.3 & 0.9 \\
       %Kim et al. 
       tb-- & \cite{park-20arXiv2012.06167} &  -0.1 & -7.4 & -0.1 & 0.05 & 0 & 1.4 \\
       %Kim et al.
       tb+ & \cite{park-20arXiv2012.06167}&  -1.5 & 3.3 & -2.8 & 2.1 & 0 & 1.5 \\
       %Samarakoon et al. 
       tc--&\cite{samarakoon-21arXiv2105.06549}$^\star$ &  -0.2 & -7 & 0.02 & -0.23 & 0.05 & 1.2 \\
       %Samarakoon et al.
        tc+& \cite{samarakoon-21arXiv2105.06549}$^\star$ & -3.2 & 2.7 & -2.9 & 1.6 & 0.1 & 1.2 \\
       td & \cite{lin_field-induced_2021}&  -2.32 & 0.125 & 0.125 & 0 & 0 & 2.5 \\
 tx-- & this work &  -0.2 & -7.0 & 0.5 & 0.15 & 0 & 1.6\\
       tx+ & this work &  -3.5 & 3.2 & -3.0 & 2 & 0 & 1.4\\
       \hline
       \hline
       \multicolumn{2}{c|}{\ncso}  & J &  K & \Gamma & \Gamma^{\prime} & J_2 & J_3\\
       \hline
       %Songvilay et al. 
       sa-- & \cite{songvilay-20prb224429} &  -2.0 & -9 & 0.3 & -0.8 & 0 & 0.8 \\ 
       %Kim et al.
       sb-- & \cite{park-20arXiv2012.06167}  & -2.1 & -4 & -0.7 & 0.6 & 0 & 1.2 \\
       %Kim et al.
       sb+ & \cite{park-20arXiv2012.06167} &  -4.7 & 3.6 & 1.3 & -1.4 & 0 & 0.95 \\
       sx--& this work & -1.4 & -10 & -0.3 & -0.6 & 0 & 0.6 \\
       sx+& this work &  -5 & 2 & -4 & 0.3 & 0 & 0.6 \\
       \hline\hline
\end{tabular}
%\end{ruledtabular}
\caption{Best fit model parameters for \ncto\ and \ncso, from present literature and our work. $\pm$
	on the labels refer to the sign of $K$.   ($\star$ This model also includes a small
interlayer Heisenberg coupling, which we neglect.) }
\label{tab:model_params}
\end{table}

\endgroup

\begin{table}

	%\begin{tabular}{|l|l|l|l|l|l|l|}
	\begin{tabular}{r|dddddd}
		\hline \hline                                                                                   
       Model  & J & K & \Gamma & \Gamma^{\prime} & J_3\\
       \hline
       tx-- & -0.20 & -7.00 & 0.50 & 0.15  & 1.40\\
       $\mathcal{T}_1$(tx+) & 0.14 & -7.73 & 0.87 & 0.18 & 1.60\\
       \hline
       tx+ &-3.50 & 3.20 & -3.00 & 2.00  & 1.60\\
       $\mathcal{T}_1$(tx--) &-3.47 & 2.80 & -2.75 & 1.78 & 1.40\\
       \hline
	sx-- & -1.40 & -10.00 & -0.30 & -0.60  & 0.60\\
	$\mathcal{T}_1$(sx+) & -2.20 & -6.40 & -1.17 & -1.10   & 0.60\\
	\hline
	sx+ & -5.00 & 2.00 & -4.00 & 0.30 & 0.60\\
	$\mathcal{T}_1$(sx--) & -5.98 & 3.73 & -4.94 & 1.69 &  0.60\\
\hline\hline
\end{tabular}
\caption{Dual pairings of best fit parameters. \label{tab:duality}}
\end{table}

%%%%%%%%%%%%%%%%%%%%%%%%%%%%%%%%%%%%%%%%%%%%%%%%%
%
%                                                 D I S C U S S I O N
%
%%%%%%%%%%%%%%%%%%%%%%%%%%%%%%%%%%%%%%%%%%%%%%%%%
\section{Discussion}
\label{sec:discussion}

\subsection{Model Duality and sign of $K$}

It is known that best fit parameters for the
extended Heisenberg-Kitaev-$\Gamma$-$\Gamma^{\prime}$ model tend to come in pairs.
 Looking at \ncto,
\ncso\ and the \ncto--analogue Na$_2$Ni$_2$TeO$_6$, 
Refs.\,\cite{park-20arXiv2012.06167, samarakoon-21arXiv2105.06549} assigned two models to each compound -- one
with a dominant $K<0$ interaction, and one with a more modest antiferromagnetic
$K$ of similar magnitude to $J_1, \Gamma$ and $\Gamma^\prime$. We will refer to these
model classes as `ferromagnetic' \,(FM, --) and `antiferromagnetic'\,(AFM, +), respectively.
Our work follows the same pattern: the top and bottom rows of Fig.\,\ref{fig:pavg} show two ostensibly unrelated models with
extremely similar powder averages.
This correspondence is due to the exact duality transformation identified by Chaloupka and
Khalliulin\,\cite{chaloupka_hidden_2015}. The $\mathcal{T}_1$ transformation as defined in their paper
rotates spin space by $\pi$ about the honeycomb plane normal vector $\mathbf{e}_{111}$,
which in our $xyz$ spin basis may be expressed as the rotation \eqref{eq:Zrot}:
\begingroup
\renewcommand{\arraystretch}{1.2}
\begin{align}
	\mathbf{S}_i &\mapsto Z \mathbf{S}_i\ , \nonumber\\[3pt]
	Z &= \begin{pmatrix}
 -\frac{1}{3} & +\frac{2}{3} & +\frac{2}{3} \\
 +\frac{2}{3} & -\frac{1}{3} & +\frac{2}{3} \\
 +\frac{2}{3} & +\frac{2}{3} & -\frac{1}{3} 
\end{pmatrix}. \label{eq:Zrot}
\end{align}
\endgroup
Note that $|Z|=1, Z = Z^T = Z^{-1}$. By applying this transformation to the $K_\gamma$ matrices
defined in Eqs.\,\eqref{Kxy} and \eqref{Kz}, one may easily verify that there exist transformed
parameters $\tilde{J},\tilde{K},\tilde{\Gamma},\tilde{\Gamma^{\prime}}$ such that
%\begingroup
%\renewcommand{\arraystretch}{1.2}
%\begin{align}
\begin{equation}
	K_\gamma(\tilde{J},\tilde{K},\tilde{\Gamma},\tilde{\Gamma^{\prime}}) =ZK_\gamma(J,K,\Gamma,\Gamma^{\prime})Z\ .
	\end{equation}
	
	This leads to a linear relationship between the original and transformed parameters.
	Following Ref.\,\cite{chaloupka_hidden_2015}, we denote this parameter transformation
	$\mathcal{T}_1$.
	Its matrix representation is presented in Eq. (\ref{eq:t1_matrix}).

	\begin{align}
	\begin{pmatrix}
		\tilde{J}\\\tilde{K}\\\tilde{\Gamma} \\ \tilde{\Gamma^{\prime}}
	\end{pmatrix}
					     &=\begin{pmatrix}
1 & +\frac{4}{9} & -\frac{4}{9} & +\frac{4}{9} \\
0 & -\frac{1}{3} & +\frac{4}{3} & -\frac{4}{3} \\
0 & +\frac{4}{9} & +\frac{5}{9} & +\frac{4}{9} \\
0 & -\frac{2}{9} & +\frac{2}{9} & +\frac{7}{9}
\end{pmatrix}
\begin{pmatrix}
J \\
K \\
\Gamma \\
\Gamma^{\prime}
\end{pmatrix}
= \mathcal{T}_1 
\begin{pmatrix}
J \\
K \\
\Gamma \\
\Gamma^{\prime}
\end{pmatrix}
\label{eq:t1_matrix}
\end{align}
%\end{align}
%\endgroup
%
Similarly, $Z^2=\mathbb{1}$ implies that any purely isotropic couplings remain fixed under $\mathcal{T}_1$, i.e.,
$\tilde{J}_i = J_i\ (i=2,3)$. In the absence of $K,\Gamma, \Gamma^\prime$, $J_1$ would also be
fixed.

This symmetry implies that two models possessing $\mathcal{T}_1$-equivalent parameters will have
indistinguishable magnon excitation spectra. Further, any choice of classical zigzag state is
site-wise symmetric under a global $\pi$ rotation about $\mathbf{e}_{111}$. This symmetry renders
the neutron scattering cross section (\ref{eq:INS cross section}) invariant under 
$\mathcal{S}^{\alpha\beta} \mapsto Z^{\alpha
\alpha'}Z^{\beta\beta'} \mathcal{S}^{\alpha'\beta'}$, implying that the two models are fundamentally
indistinguishable by scattering experiments that couple only to the spins. The INS cross section is given by
\begin{equation}\label{eq:INS cross section}
\frac{d^2\sigma}{d\Omega dE_f} \propto F(Q)^2 S^\perp(\mathbf{Q},\omega)
\end{equation}
with the dynamical structure factor
\begin{equation}\label{eq:sperp}
	S^\perp(\mathbf{Q},\omega) = \sum_{\alpha\beta} \left(\delta_{\alpha\beta} - \frac{Q^\alpha Q^\beta}{Q^2}
	\right)
	\mathcal{S}^{\alpha\beta}(\mathbf{Q}, \omega) 
	\end{equation}
	and its matrix elements
	\begin{equation}	
		\mathcal{S}^{\alpha\beta}(\mathbf{Q},\omega) = \sum_{ij} \int \frac{d\tau}{2\pi} e^{-i\omega \tau} \left\langle
			S^\alpha_{-\mathbf{Q}i}(0)
S^\beta_{\mathbf{Q}j}(\tau) \right\rangle\ .
\end{equation}
The powder structure factor is obtained by averaging over all momentum transfer directions.
\begin{equation}\label{eq:INS cross section avg}
\int d\Omega \frac{d^2\sigma}{d\Omega dE_f} \propto F(Q)^2 \int d\Omega S^\perp(\mathbf{Q},\omega)
\end{equation}
The four inequivalent sites in the zig-zag ground state's magnetic unit cell are indexed by $i, j$, 
$F(Q)$ is a spherically symmetric atomic scattering factor, and $S^\alpha_{\mathbf{Q}j}(\tau)$ is a
Fourier mode of a Heisenberg picture spin operator. We neglect the Debye-Waller factor due to the
low temperature.

The symmetry pertains to the spin model, not the linearised spin wave approximation. The two
parameter sets are
therefore \textit{functionally} indistinguishable -- the antiferromagnetic (+) parameter sets,
which would seem to have their weak $K$ parameters suppressed by large $J_{1,2,3}$, $\Gamma, \Gamma'$ terms,
have equivalent excitation spectra to models with dominant $K<0$. Note that there are \textit{two}
axes in $J_1,K,\Gamma,\Gamma'$ space corresponding to pure Kitaev models: the obvious $(0,K,0,0)$
axis, and the axis generated by $\mathcal{T}_1(0,K,0,0) = K(\frac{4}{9}, -\frac{1}{3}, \frac{4}{9},
-\frac{2}{9})$ for any $K\neq 0$. In general, when off-diagonal contributions are sufficiently small
a large $K<0$ parameter is mapped to a mixture of $K, \Gamma,
\Gamma^\prime$ with $K$ taking the opposite sign. 
Tab.\,\ref{tab:duality} confirms that this is essentially the case with our model pairs, and indeed
the visualisation in Fig.\,\ref{fig:spinorient} confirms that the same is true of most other
published models.

Arguably, this makes both parameter sets equally `close' to a
Kitaev spin liquid. All models presented in Tab. \ref{tab:model_params} (with the notable exceptions
of td and sb+) either have dominant $K<0$, or are $\mathcal{T}_1$-equivalent to a model with dominant
$K$, \ie close to the Kitaev point \textbf{K}-- (Fig.\,\ref{fig:spinorient}) which is discussed further below.
\ncso\ and \ncto\ are therefore good candidate materials for field-revealed or strain-revealed quantum spin
liquids.

The Hubbard models studied in Refs.\,\cite{liu-20prl047201, liu_pseudospin_2018,
sano_kitaev-heisenberg_2018} indicate a robustly ferromagnetic Kitaev
interaction over a wide range of parameters.
The AFM models arguably all suffer from fine tuning in the $\Gamma, \Gamma'$ parameters; a change
of only a few percent in either parameter opens the gap to many meV.
The FM models correspond to a
small trigonal-field perturbation of an ideal edge-sharing geometry, and so would seem to be the
more likely explanations of scattering in \ncto\ and \ncso.

%%%
%%%
%%%
\subsection{Existence of a gap}

Powder spectra from \ncto\ [Fig.\,\ref{fig:ins_data}\,(a), (c), (e)] exhibit a clear spin-wave mode emanating from
$Q=0.75$ \AA$^{-1}$, corresponding to the $M$ point
of the in-plane Brillouin zone. Our data show a weak spectral weight extending all the way down to the
elastic line, which may in principle be evidence of a gapless magnon excitation. However, our
best-fit results agree with other papers that the spectrum is gapped, though we revise the size of
this gap down to  1meV$\pm0.3$meV.

\begin{figure}[t!]
\centering
\includegraphics[width=\columnwidth]{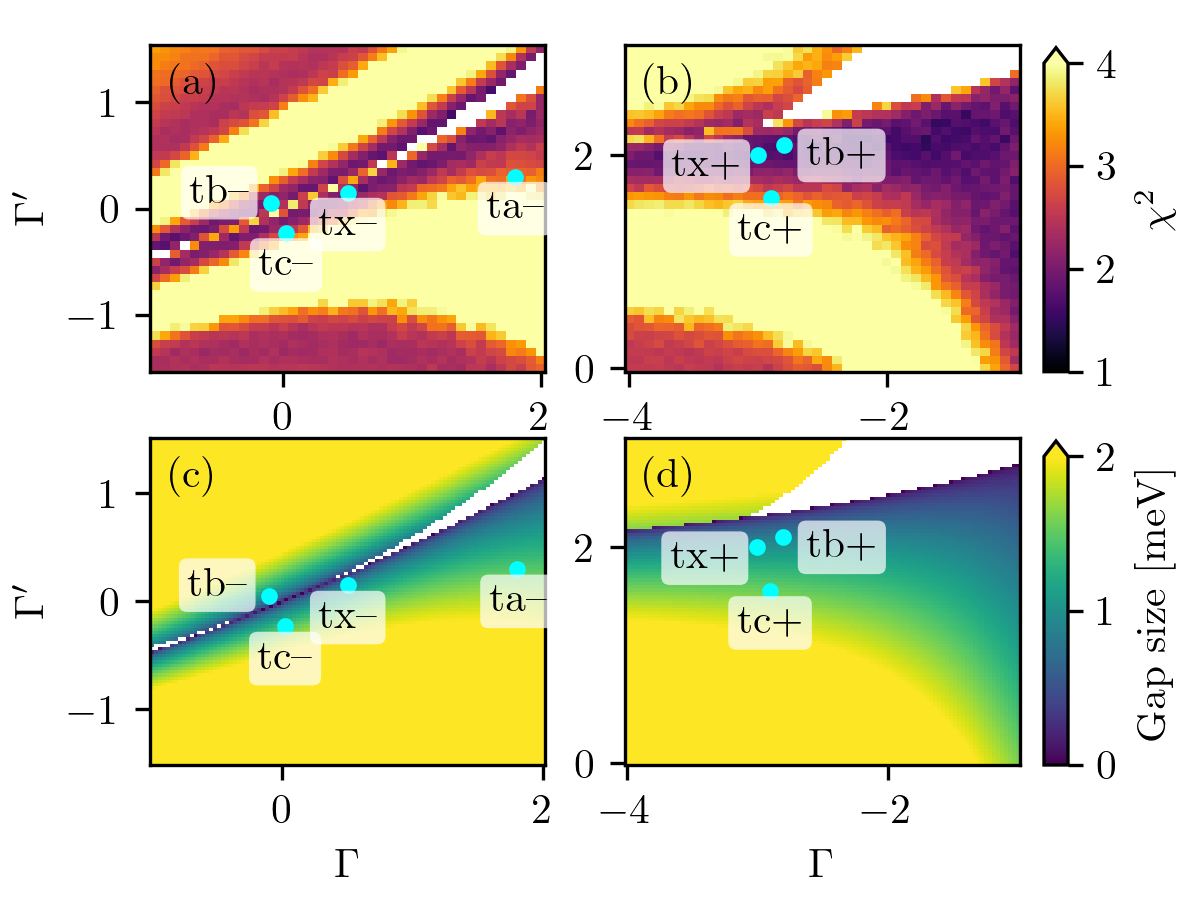}
\includegraphics[width=\columnwidth]{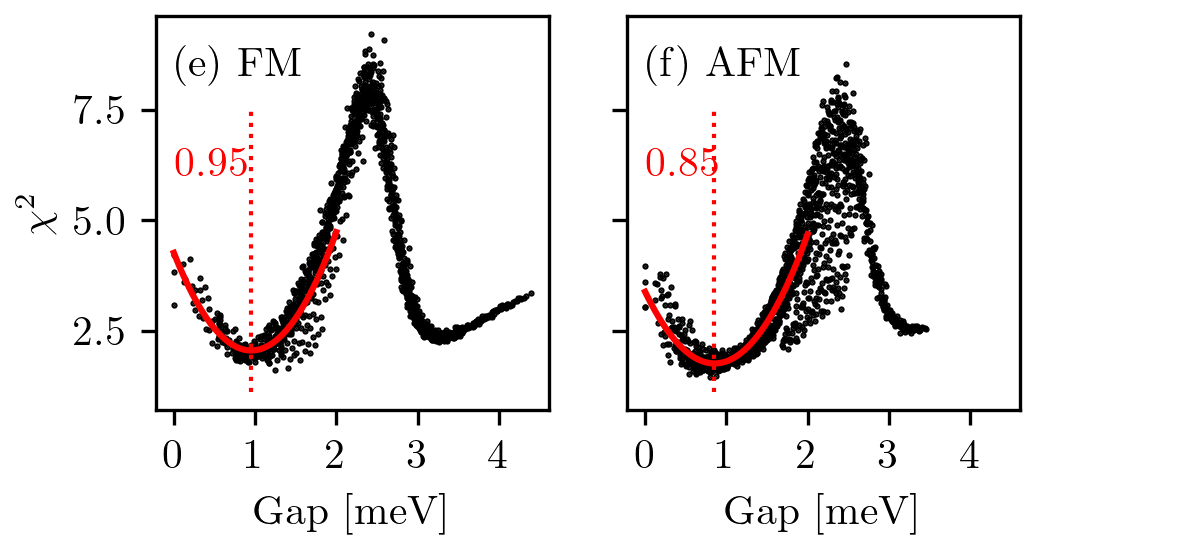}
\caption{$\chi^2$ values and gap sizes for perturbations of the tx- (left column) and tx+ (right
column) \ncto\ models in $\Gamma-\Gamma^\prime$ space, both of which are in units of meV. White regions indicate that the Hamiltonian is
not positive definite, meaning that the classical magnetic structure is not zig-zag. Pixel size is
indicative of grid used for parameter generation. All $\chi^2$ values from (a),(b) are
shown against their corresponding gap in (e), (f), with interpolating parabolas as a guide to the eye. } 
	\label{fig:gap}
\end{figure}

Our model's $M$ point gap is controlled primarily by the parameters $\Gamma$ and $\Gamma'$. 
As can be seen in Fig.\,\ref{fig:gap}(a) and
(b), the goodness of fit has a W-shaped valley on either side of the gapless line.
Crucially, the truly gapless models correspond to a local maximum while the best fit values of
$\Gamma, \Gamma^\prime$ for all published models listed here have gaps of order ~1 meV. 
The plot of $\chi^2$
	against gap size [Fig. \ref{fig:gap}\,(e),(f)] shows a very clear minimum near 0.9meV. One may
	heuristically assign an uncertainty of 0.3meV based on the spread of gap sizes with comparable
	goodness of fit.

We take this to be strong evidence that \ncto\  has a gapped spin-wave
spectrum. We attribute the weak excess of spectral weight below 1 meV in the low temperature data
[Fig.\,\ref{fig:ins_data}\,(c)] to energy broadening from
magnon decoherence. The material's powder averaged magnon spectrum therefore bears a very close resemblance to that of
$\alpha$-RuCl$_3$ at zero magnetic field\,\cite{banerjee_proximate_2016,banerjee_excitations_2018}.

The 50mK powder spectrum of \ncso\ [Fig.\,\ref{fig:ins_data}\,(b), (d)] tells a different story. There is no evidence of a spin
wave mode emanating from the $M$ point, all spectral weight is concentrated near a broad feature at the
$\Gamma$ point. Both best-fit models concur that the mode is quadratic
with a gap of order $\sim1$ meV, as can be seen in the powder averaged theory predictions [Fig.
\ref{fig:pavg}\,(b),\,(d)]. A more precise bound is not possible due to the occlusion of the
$\Gamma$ point by
the zero-order neutron peak.

%%%%%%%%%%%%%%%%%%%%%%%%%%%%%%%%%%%%%%%%%%%%%%%%%
%%%%%%%%%%%%%%%%%%%%%%%%%%%%%%%%%%%%%%%%%%%%%%%%%

Fig.\,\ref{fig:ins_data}\,(e) shows the appearance of an unexpected `hourglass' feature when the
temperature is increased from 50\,mK. The excitation near zero energy is very broad in $Q$, and has a clearly observable waist
that excludes the possibility that it is the result of magnon lifetime broadening (see the
finite-temperature modelling of $\alpha$-RuCl$_3$ modelled in Refs.
\cite{maksimovRethinkingEnsuremathAlpha2020, smitMagnonDampingZigzag2020,
winterBreakdownMagnonsStrongly2017}).
The feature we see is shadowed by elastic scattering in many experiments
\cite{songvilay-20prb224429, samarakoon-21arXiv2105.06549}, but is arguably visible (albeit in low
resolution) in Fig. 5 of
Ref. \cite{park-20arXiv2012.06167}, taken at 3K. The feature is not readily discernible in the single-crystal
data of \cite{chen-21prbL180404}, also at 3K.

Though it is tempting to draw comparisons with the hourglass magnetic excitations of copper oxide
superconductor relatives \cite{boothroydHourglassMagneticSpectrum2011,
dreesHourglassMagneticExcitations2014}, 
it is more likely that the signal we observe arises from broadening of the
magnetic Bragg peak in energy-momentum space, which is consistent with a finite average domain size
\cite{kataninQuasielasticNeutronScattering2011a}.
The apparent absence of this signal in the single crystal data may be attributed to its diffuse
structure - by their very nature, single-crystal experiments sample only a small
volume of energy-momentum space.

%%%%%%%%%%%%%%%%%%%%%%%%%%%%%%%%%%%%%%%%%%%%%%%%%
\begin{figure}[t!]
	\centering
	\includegraphics[width=\columnwidth]{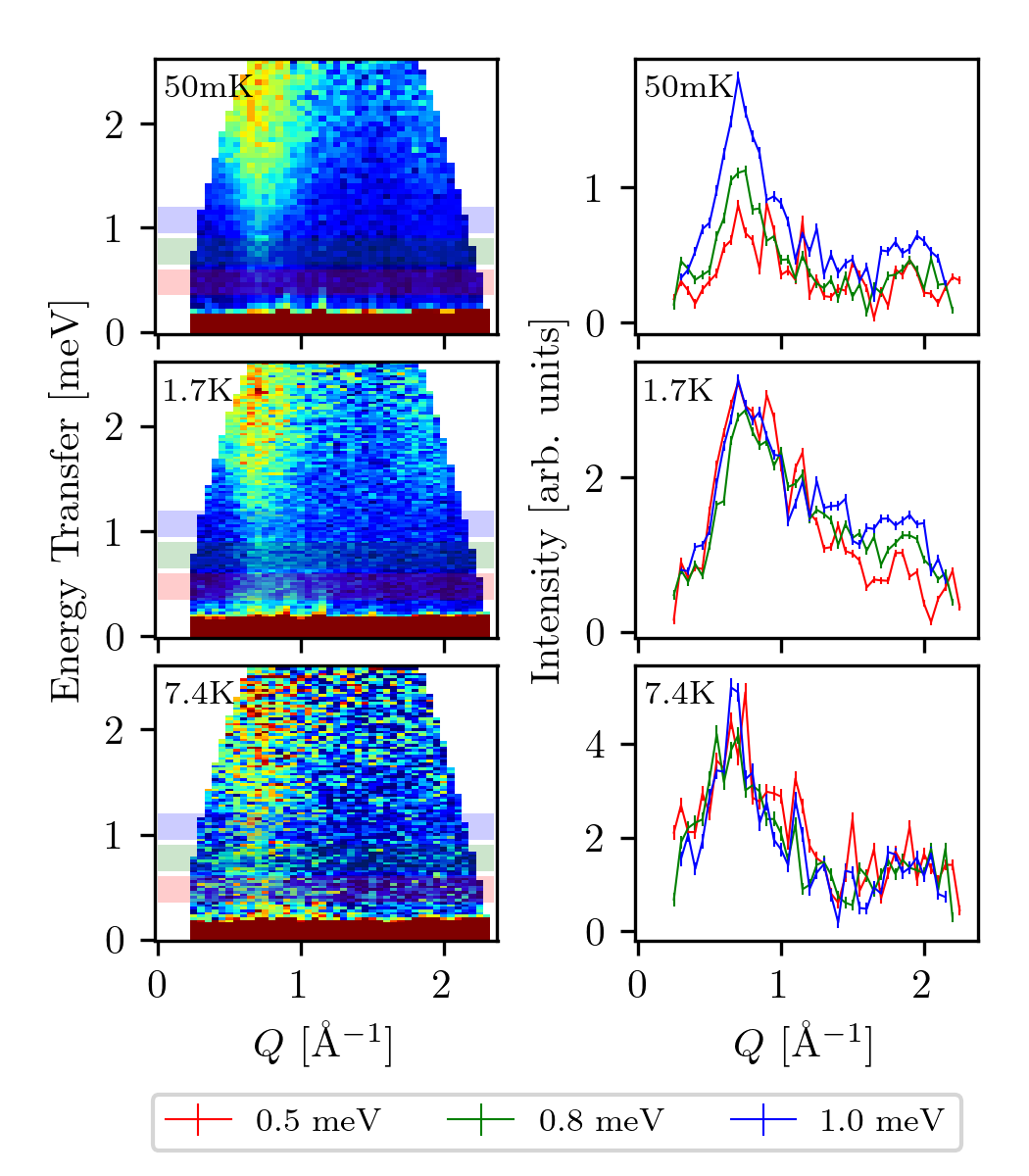}
	\caption{Constant energy slices revealing the `hourglass' dispersion close to zero energy at the
	$M$ point of \ncto\ for $T=0.05$\,K (top row), $T=1.7$\,K (middle row) and $T=7.4$\,K (bottom row).
}
	\label{fig:hourglass}
\end{figure}
%%%%%%%%%%%%%%%%%%%%%%%%%%%%%%%%%%%%%%%%%%%%%%%%%

\subsection{Classical ground state}
\label{sec:cgs}

\begin{figure*}[]
	\centering
	\includegraphics[width=\textwidth]{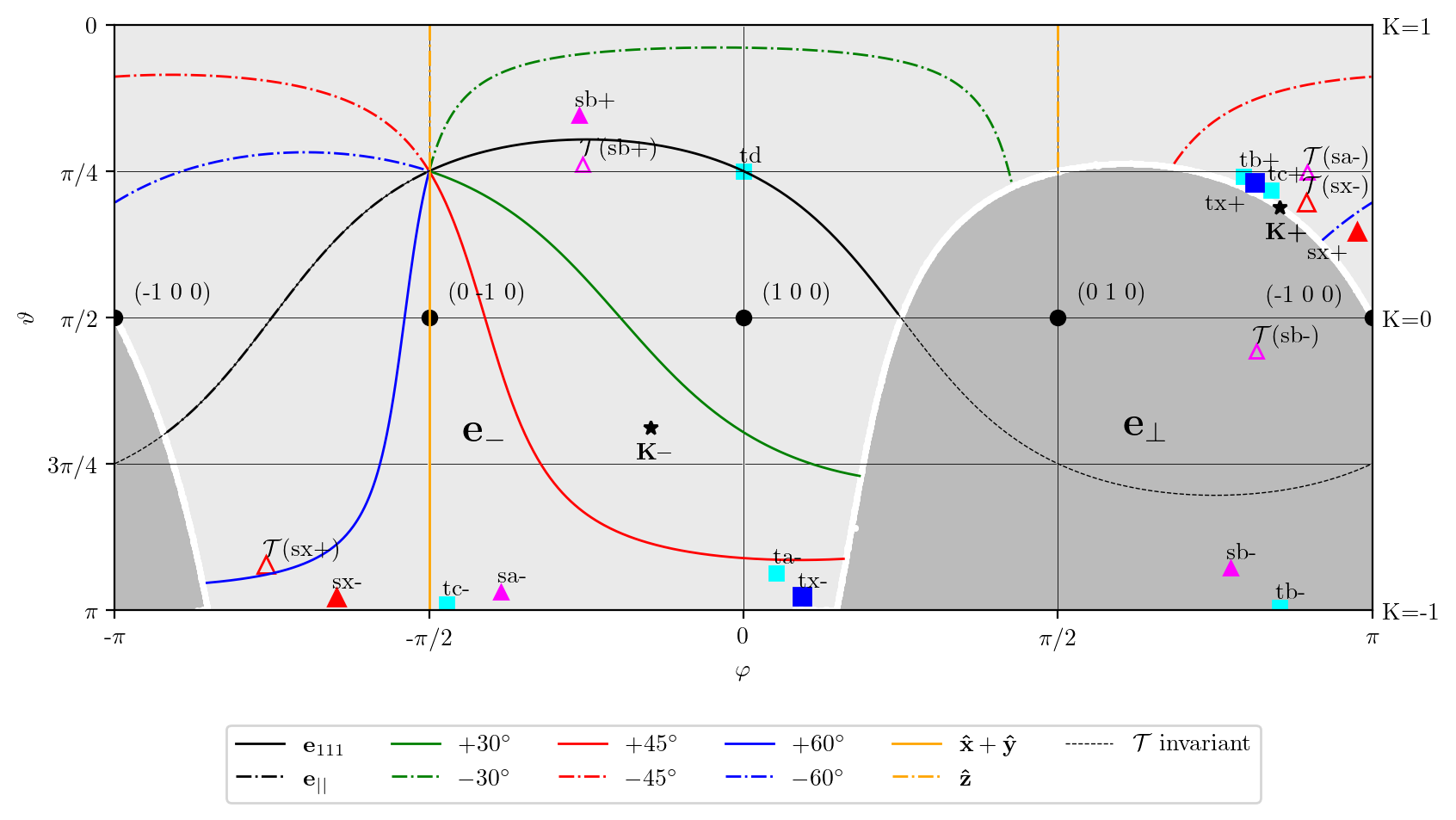}
	\caption{
		Ground-state zig-zag magnetic moment orientations over the sphere $\Gamma^2 + (\Gamma^{\prime})^2 +K^2 = 1$ in
		$(\Gamma, \Gamma^{\prime}, K)$ space, assuming zigzag ground state stability. The sphere is parameterised using $\Gamma = \sin(\vartheta)\cos(\varphi), \Gamma^{\prime} = \sin(\vartheta)\sin(\varphi), K =
	\cos(\vartheta)$. 
	In the $\mathbf{e}_-$ phase, meridians are labeled by the angle of $\mathbf{e}_-$ to
	$\mathbf{e}_{111}$. The fine black dotted line is the intersection of the
	$\mathcal{T}_1$-invariant plane $K-\Gamma+\Gamma^\prime=0$ with the unit sphere. \textbf{K+}
	and \textbf{K--}
	indicate the nontrivial `hidden' Kitaev points\,\cite{chaloupka_hidden_2015} generated by $\mathbf{K}\pm =
	\mathcal{T}_1(0,0,\mp1)$. 
	Models are defined in Tab.\,\ref{tab:model_params}.
}
\label{fig:spinorient}
\end{figure*}

We make the ansatz that the material is in a colinear zig-zag phase, featuring ferromagnetic chains oriented normal to the
$z$-bond direction. The classical magnetic ground state vector
corresponds to the eigenvector
of the matrix $M := K_x + K_y - K_z$ of smallest eigenvalue\,\cite{park-20arXiv2012.06167}. The Heisenberg interactions shift all eigenvalues
uniformly, with no effect on the eigenvector: the eigenvalues and eigenvectors depend only on $K,
\Gamma$, and $\Gamma^\prime$, with the explicit forms \cite{park-20arXiv2012.06167}
\begin{align}
\nonumber	\mathbf{e}_\perp &\propto(-1,1,0)\ ,\\[3pt]
\nonumber	\mathbf{e}_\pm &\propto
	(2x -1 \pm \zeta, 2x -1 \pm \zeta, 4) & (\Gamma \neq 0)\\[3pt]
\nonumber	\lambda_\perp &= \Gamma + K - 2\Gamma^{\prime}\ ,\\[3pt]
\nonumber	\lambda_\pm &= \left( 2 \Gamma^{\prime} - \Gamma \pm |\Gamma\zeta|\right)/2\ .
\end{align}
It is convenient to define $\mathbf{e}_\pm$ by analytic continuation when $\Gamma=0$, excepting the pure-Kitaev
special cases which correspond to a triple degenerate hidden SO(3) symmetry.
For convenience, we also define the basis vectors from Fig.\,\ref{fig:3D},
\begin{align}
	\mathbf{e}_\| &= \frac{1}{\sqrt{6}}(1,1,-2)\ ,\nonumber \\
	\mathbf{e}_{111} &= \frac{1}{\sqrt{3}}(1,1,1)\ . \nonumber
\end{align}
We have set $\zeta = \operatorname{sgn}(\Gamma)\sqrt{8 + (2x-1)^2}$ and $x = (K+\Gamma^{\prime}) / \Gamma$. Note that $\lambda_+
> \lambda_-$. 
When $x = 1$, $\mathbf{e}_\pm$ are aligned with
$\mathbf{e}_\|$ and $\mathbf{e}_{111}$. Changes in $x$ rotate this orthogonal
eigenbasis about $\mathbf{e}_\perp$. The signed rotation angles for various $\Gamma,\Gamma',K$ are
shown superimposed on Fig.
\ref{fig:spinorient}.

%%%%%%%%%%%%%%%%%%%%%%%%%%%%%%%%%%%%%%%%%%%%%%%%%%%
%%%%%%%%%%%%%%%%%%%%%%%%%%%%%%%%%%%%%%%%%%%%%%%%%%%

Given that
a minimal eigenvector remains a minimal eigenvector under the uniform rescaling 
$M \mapsto \alpha M, \alpha \in \mathbb{R}_{>0}$, \ie the zig-zag orientation is not sensitive to
the overall energy scale, it is
possible to represent all possible spin alignments of zig-zag ground states by looking at the unit sphere in
$\Gamma-\Gamma^{\prime}-K$ space.
The action of $\mathcal{T}_1$ on the theory space can be visualised as mirroring about the plane
$K+\Gamma^\prime = \Gamma$. On this plane, the spin orientations must be one of $\mathbf{e}_{111},
\mathbf{e}_{||}$ and $\mathbf{e}_\perp$, which are clearly
eigenvectors of $Z$ with eigenvalues $1, -1$ and $-1$ respectively.

Crucially, the two pure Kitaev points at the north and south poles are inequivalent.
This is true even for a classical spin model; $K=-1$ is a degeneracy point between $\mathbf{e}_-$
and $\mathbf{e}_\perp$, $K=1$ is deep in the $\mathbf{e}_-$ region.
Under $\mathcal{T}_1$ they are mapped not to each other but to other points in parameter space,
marked \textbf{K}$\pm$, which we call `hidden' Kitaev points.
The cluster of AFM \ncto\ models near
\textbf{K--} is (approximately) mapped to the cluster near the $K=-1$ pole and vice verse, as
are the sx$\pm$ models (see Fig.\,\ref{fig:spinorient}).

Representation-theoretic analyses of X-ray and neutron diffraction data suggest that the spins
in \ncto\ are aligned close to $\mathbf{e}_\parallel$\,\cite{bera_zigzag_2017, lefrancois_magnetic_2016}, with angular uncertainty of order $\sim
30^\circ$. The \ncto\ models are only slightly beyond this uncertainty. It should be emphasised that
since these models are found near the critical triple-degenerate Kitaev point at $(0,0,-1)$ in the $(\Gamma, \Gamma', K)$ basis, quantum
fluctuations are expected to be large, and higher order magnon interactions may tune the effective zig-zag
direction as seen by elastic scattering. 
For tx--, the classical energy difference between the $\mathbf{e}_\perp$ and $\mathbf{e}_-$
phase is of order $\delta E =$ 0.4\,meV.
Working so close to a classical degeneracy, it is to be expected that the classical ground state
calculation will be somewhat inaccurate.
The established cluster of models ta--, tb$\pm$, tc$\pm$ are therefore loosely compatible with the reported structure.
We note in passing that the anomalous td model from Ref.\,\cite{park-20arXiv2012.06167} is a special case,
occurring on the nodal $\mathcal{T}$ invariant line with a ground state along $\mathbf{e}_{111}$,
which is in tension with the experimentally ascertained magnetic order.

\subsection{Prediction of single crystal results}

%%%%%%%%%%%%%%%%%%%%%%%%%%%%%%%%%%%%%%%%%%%%%%%%%%%%%%%%%%%
\begin{figure}[h!]
\centering
\includegraphics[width=\columnwidth]{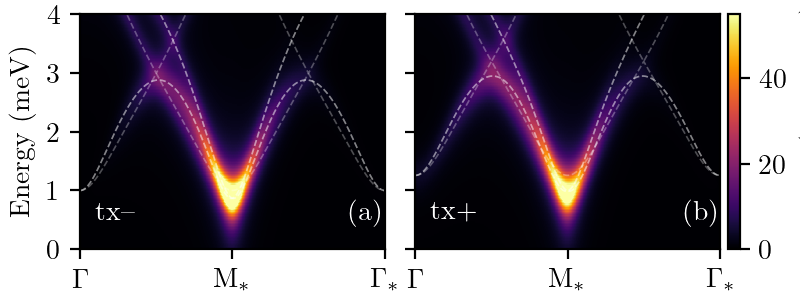}
\includegraphics[width=0.3\columnwidth]{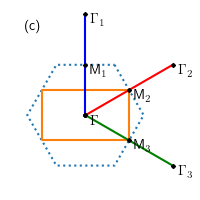}
\includegraphics[width=0.65\columnwidth]{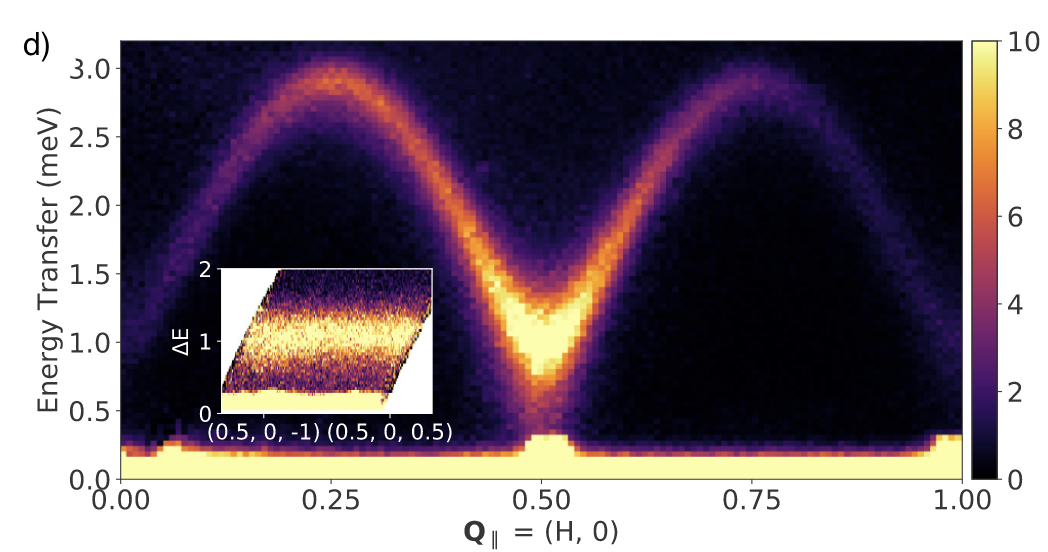}
\includegraphics[width=\columnwidth]{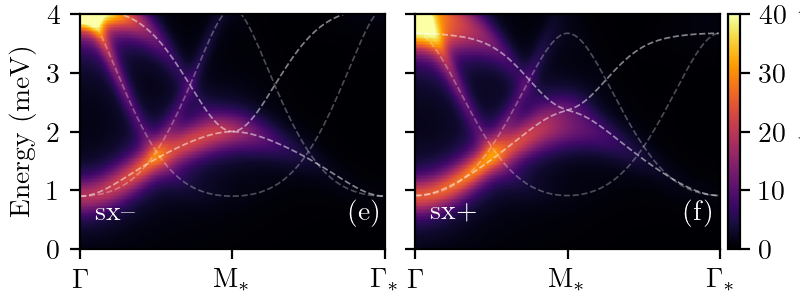}
\caption{Domain average over the three Brillouin zone walks (red, green, blue) sketched in (c) for 
	(a) FM and (b) AFM models of \ncto. The white dashed lines correspond to the LSWT dispersion relation.
(d) Single crystal experimental \ncto\ data from Ref.\,\cite{chen-21prbL180404} for comparison. (e, f)
show analogous plots to (a, b) for \ncso. Linear magnon dispersions are shown in white. $\bs{Q}_{||}=(0,0)$ corresponds to the $\Gamma$ point, and $\bs{Q}_{||}=(1,0)$ to $\Gamma^{\prime}$ and the strong accumulation of spectral weight thus appers at $M$.}                                                                       
\label{fig:singlecrystal}
\end{figure}
%%%%%%%%%%%%%%%%%%%%%%%%%%%%%%%%%%%%%%%%%%%%%%%%%%%%%%%%%%%

Fig.\,\ref{fig:singlecrystal} shows a comparison between domain averages of the \ncto\ models tx+ and tx-- and the single crystal data of
Ref.\,\cite{chen-21prbL180404}. A single crystal will in general consist of a statistical population of three $C_3$ related domains
corresponding to $x$, $y$ and $z$ zig-zags. An experiment will therefore detect an average over the
three $k$ space paths shown in Fig.\,\ref{fig:singlecrystal}\,(c). The paths $\Gamma \to \Gamma_2$ and
$\Gamma \to \Gamma_3$ are equivalent under the $D_2$ symmetry of the Brillouin zone, implying that
an experiment should always detect an even number of modes. However, this is not seen in the
experimental data, leading the authors of Ref.\,\cite{chen-21prbL180404} to suggest that \ncto\
undergoes a spontaneous charge transfer that renders 1/4 Co sites spinless, leaving the remainder to
follows a vortex-like triple-$Q$ order. 
However, our calculations in Fig.\,\ref{fig:singlecrystal}\,(a) and (b) show that magnon modes arising from an ensemble of zig-zag orders are equally capable of reproducing the dispersion
in Fig.\,\ref{fig:singlecrystal}\,(d).
Each domain contributes two low-energy magnon modes, resulting in four inequivalent modes. Notably, the doubled modes are the
sinusoid-like objects matching the experiment. The missing modes are suppressed by the spin
orientation, leading to vanishing spectral weight. 
While the overall agreement  between the single crystal data of Ref.\,\cite{chen-21prbL180404} and
our theory modeling fitted to our powder sample is surprisingly good, the theoretical prediction is
not an exact match. The regions $Q_{\|}<0.25$, $Q_{\|} > 0.75$ appear
too weak in the theoretical prediction, while the crossover at $Q_{\|}=0.25$ is absent in the
experiment. The calculations here are done in the linear approximation, so
slight deviations when far from the vacuum are expected. In particular, the modes near high density
of states at the $M$ point and the crossover would be expected to have stronger magnon damping, suppressing their INS signals.

The predicted single-crystal scattering signals for the sx-- and sx+ \ncso\ models in Fig.\,\ref{fig:singlecrystal}\,(e) and \ref{fig:singlecrystal}\,(f) are more easily distinguished from each
other than the tx$\pm$ models. This is a consequence of the lack of distinguishing features in the
\ncso\ powder spectrum, which in turn hinders the accurate determination of best-fit parameters. The
dispersion ranges over an energy scale of several meV well above the elastic line,
with substantial spectral weight away from the $\Gamma$ point. 
It is therefore feasible that these features may be observed in a future single
crystal experiment on \ncso.

%%%%%%%%%%%%%%%%%%%%%%%%%%%%%%%%%%%%%%%%%%%%%%%%%
%
%                                                C O N C L U S I O N
%
%%%%%%%%%%%%%%%%%%%%%%%%%%%%%%%%%%%%%%%%%%%%%%%%%
\section{Conclusion}
\label{sec:conclusion}

In this paper we have presented results from INS for \ncso\ and \ncto\  measured down to 50\,mK. Analysis of the data and comparison with our theoretical modeling shows that spin waves arising from a zig-zag ground state of an
extended Heisenberg--Kitaev model are capable of
representing the INS behaviour of \ncto\ and \ncso. The fine detail that our experiment provides of
the $M$-point excitations of \ncto\ has allowed us to refine the existing cluster of parameters.

The failure of naive Heisenberg and XXZ models to recreate the magnon dispersions of these materials\,\cite{songvilay-20prb224429} should be taken as strong evidence of the importance of anisotropic
spin couplings. In particular, our \ncso\ data show a gapped spectrum with no evidence of linear
modes emanating from the $M$ point, which is suggestive of substantial SO(3) symmetry breaking.

We have searched for antiferromagnetic and ferromagnetic best-fit models of \ncto\ and \ncso,
and demonstrated that they are mapped to each other under the exact duality transformation
$\mathcal{T}_1$ of Ref.\,\cite{chaloupka_hidden_2015}. 
This gives us confidence that our model fitting procedure has converged to a global
minimum, or at least as close to it as experimental resolution allows. 

Fine detail of the $M$ point of \ncto\ gives strong evidence for the existence
of a small gap, which we conservatively estimate to be 1.0(5)\,meV. This gap is consistent with the
dispersion measured by the single-crystal experiment of Ref.\,\cite{chen-21prbL180404}, and the
parameters obtained are close to the existing cluster.

Our \ncso\ data show an energy minimum at the $\Gamma$ point,
however practical constraints prevent the direct observation of the $Q=0$ gap. 
Our use of 4.69 \AA\  neutrons
has given us fine detail as close as reasonably practicable to the $\Gamma$ point, and our
best fit extrapolations remain consistent with a
broad, quadratic mode. This implies substantial non-spontaneous breaking of SU(2) spin symmetry, however the lack
of clearly resolvable features in the spectrum means that we cannot categorically exclude models
other than the work presented here. The calculations in Fig.\,\ref{fig:singlecrystal}\,(e) and (f) show that
the predicted magnon dispersion along the $\Gamma \to \Gamma_*$ line is clearly observable in a single crystal INS experiment.

Our models of both substances [Tab. \ref{tab:duality}] emphasise the dominance of $K$ over other parameters,
suggesting that these materials are very close to the Kitaev spin liquid phase. This would validate
the \textit{ab initio} arguments about the nature of 3d$^7$ materials as opposed to the d$^5$ metals
Ir and Ru that have dominated searches in the past
decade\,\cite{winter_models_2017,banerjee_proximate_2016,choi-12prl127204,janssenMagnetizationProcessesZigzag2017}, specifically that the tighter orbital confinement
suppresses the direct Co-Co exchange responsible for Heisenberg interactions
\cite{liu_pseudospin_2018, liu_towards_2021, kimSpinorbitalEntangledState2021}. These models are
consistent with \textit{ab initio} calculations, which unambiguously show that cobaltate honeycombs have a
ferromagnetic $K$. We consider the
AFM models to be the less likely set of parameters -- they have substantial off-diagonal couplings,
corresponding to large trigonal distortions of the honeycomb lattice, and are in some sense fine-tuned.

Current INS experimental data cannot distinguish between AFM and FM models linked by $\mathcal{T}_1$
duality. This ambiguity has recently been conclusively resolved in $\alpha$-RuCl$_3$ by the use of
resonant elastic X-ray scattering on a high-quality single crystal
\cite{searsFerromagneticKitaevInteraction2020}. In principle, INS with a symmetry breaking in--plane
magnetic field might be an alternative strategy to distinguish the two.

Our best fit models of \ncto\ and \ncso\ seem to suggest that the substitution of tellurium by antimony
only marginally affects the spin exchange physics. Recent experiments have observed that zig-zag order breaks down
for \ncto\ at large magnetic fields, potentially entering a spin-liquid
state\,\cite{lin_field-induced_2021, hong-21arXiv2101.12199}. To the authors' best knowledge, \ncso\ has not been probed with a
magnetic field to date. Our results suggest the cross-hexamer $J_3$ coupling, largely responsible
for stabilising the zig-zag state, may be weaker in \ncso, suggesting a possibly lower critical magnetic field.
We therefore suggest that our work be read as further motivation for the study of \ncso\ as a
field-revealed Kitaev QSL candidate.

\acknowledgements
The authors acknowledge discussions with M.\ Vojta, L.\ Janssen, I.\ I.\ Mazin and H.\ O.\ Jeschke.
S.R.\ and C.D.L. acknowledge support from the Australian Research Council through Grants No.\ FT180100211 and DP200100959, respectively.
We also acknowledge the provision of beamtime (under proposal P7327) by the Australian Centre for
Neutron Scattering and support from the sample environment team. 

\bibliography{cobaltates}

%\FloatBarrier

%\newpage
%\phantom{x}
%\newpage

\appendix

\section{Details of \ncto\ broadening function determination}
\label{sec:appA}

We determined the Lorentzian lifetime broadening phenomenologically, by performing a Voigt profile
fit to the availalbe single crystal data.

\begin{figure}[h!]
	\centering
	\includegraphics[width=\columnwidth]{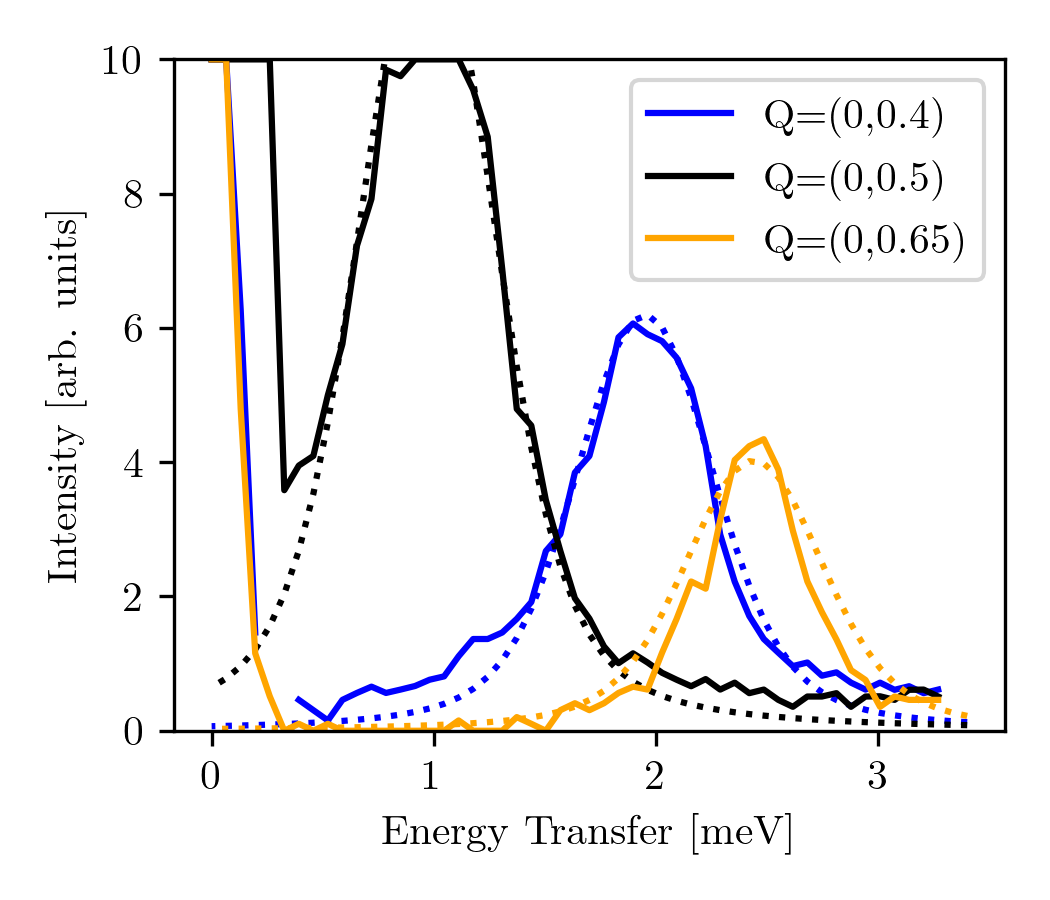}
	\caption{Empirical fit of slices from single crystal data in \cite{chen-21prbL180404} by Voigt functions, each with $\gamma =
		0.25$\,meV and $\sigma = 0.135$\,meV. Note that the $M$ point feature (notated here as
		$Q=(0,0.5)$ reciprocal lattice units) shows the strongest broadening, while the higher energy features
		are sharper.
	}
	\label{Sfig:fits}
\end{figure}

\end{document}